\documentclass[twocolumn,twoside,slac_two]{revtex4}
\usepackage{graphicx}
\usepackage{fancyhdr}
\newcommand{\pks}{PKS\,0625$-$354 }
\newcommand{\tc}{3C\,78}

\begin{document}

\title{Suzaku and Fermi Observations of Gamma-Ray Bright Radio Galaxies: Origin of the X-ray Emission and Broad-Band Modeling}

%

\author{Y. Fukazawa, S. Tokuda, R. Itoh, Y. Tanaka}
\affiliation{Department of Physical Science, Hiroshima University; Hiroshima Astrophysical Science Center, Hiroshima University}
\author{J. Finke}
\affiliation{U.S.\ Naval Research Laboratory}
\author{L. Stawarz}
\affiliation{Institute of Space and Astronautical Science, JAXA; Astronomical Observatory, Jagiellonian University}

\begin{abstract}
We performed a systematic X-ray study of eight nearby $\gamma$-ray
 bright radio galaxies with {\em Suzaku} for understanding the
 origin of their X-ray emissions. The {\em Suzaku} spectra for
 five of those have been presented previously, while the remaining three
 (M\,87, PKS\,0625$-$354, and 3C\,78) are presented here for the first
 time. Based on the Fe-K line strength, X-ray variability, and X-ray
 power-law photon indices, and using additional information on the [O
 III] line emission, we argue for a jet origin of the observed X-ray
 emission in these three sources.  We also analyzed five years of {\em
 Fermi} Large Area Telescope (LAT) GeV gamma-ray data on PKS\,0625$-$354
 and 3C\,78 to understand these sources within the blazar picture. 
We found significant 
$\gamma$-ray variability in the
 former object.  Overall, we note that the {\em Suzaku} spectra for both
 PKS\,0625$-$354 and 3C\,78 are rather soft, while the LAT spectra are
 unusually hard when compared with other $\gamma$-ray detected low-power
 (FR\,I) radio galaxies. We demonstrate that the constructed broad-band
 spectral energy distributions of PKS\,0625$-$354 and 3C\,78 are well
 described by a one-zone synchrotron/synchrotron self-Compton model.
 The results of the modeling indicate lower bulk Lorentz factors
 compared to those typically found in other BL Lac objects, but
 consistent with the values inferred from modeling other LAT-detected
 FR\,I radio galaxies. Interestingly, 
the modeling also implies very high peak ($\sim 10^{16}$\,Hz)
 synchrotron frequencies in the two analyzed sources, 
contrary to previously-suggested scenarios for FR I/BL Lac unification. 
We discuss the implications of our findings in the context of 
the FR\,I/BL Lac unification schemes.

\end{abstract}

\maketitle

\thispagestyle{fancy}


\section{Introduction}

This contribution is based on \citet{fuk14}. Here, we
very briefly describe the digest.
{\em Fermi} Large Area Telescope (LAT) established that radio galaxies
are bright gamma-ray emittors \citep{abd10b}.
However, inner jet emission has been detected mainly in the radio and
GeV gamma-ray band for most object, due to bright stellar and
accretion disk components in the optical and X-ray band;
Spectral Energy Distribution (SED) of jet emission is often
unclear, even for Cen\,A and NGC\,1275.
Thus, X-ray detection of jet emission is important for SED modeling.

{\em Suzaku} X-ray satellite has observed 8 nearby GeV-emitting radio
galaxies listed in \citet{abd10b}; some of observations are originally
proposed by ourselves.
Most of {\em Suzaku} results has been published, and some of them
exhibit a Fe-K line (3C\,111, 3C\,120, NGC\,1275, Cen\,A) and 
others do not (NGC\,6251, M\,87).
Here we report {\em Suzaku} and {\em Fermi} results on \pks and \tc 
\citep{fuk14}.

\section{X-ray Results}

\begin{figure}[!t]
\begin{center}
\includegraphics[scale=0.32]{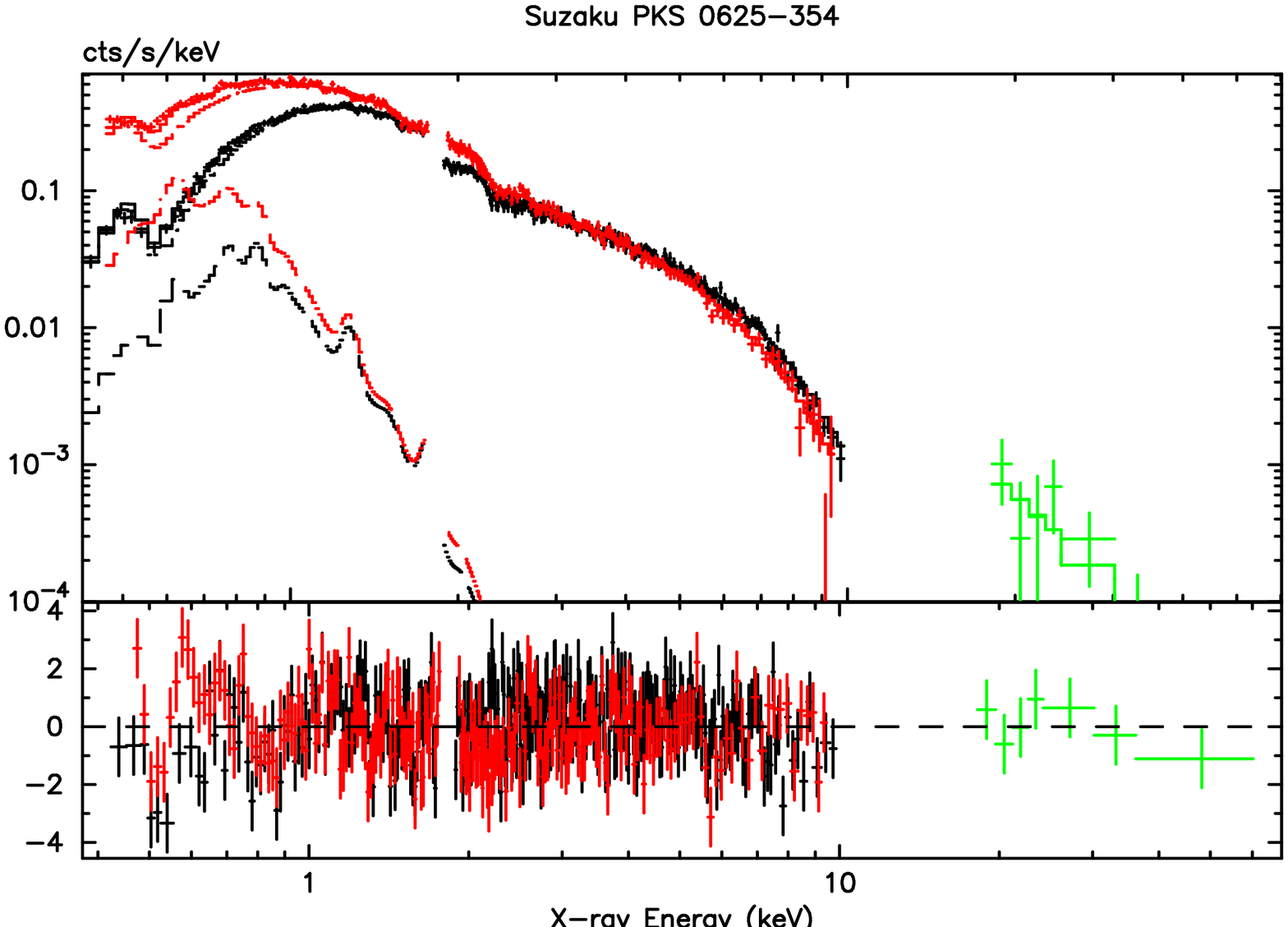}
\includegraphics[scale=0.32]{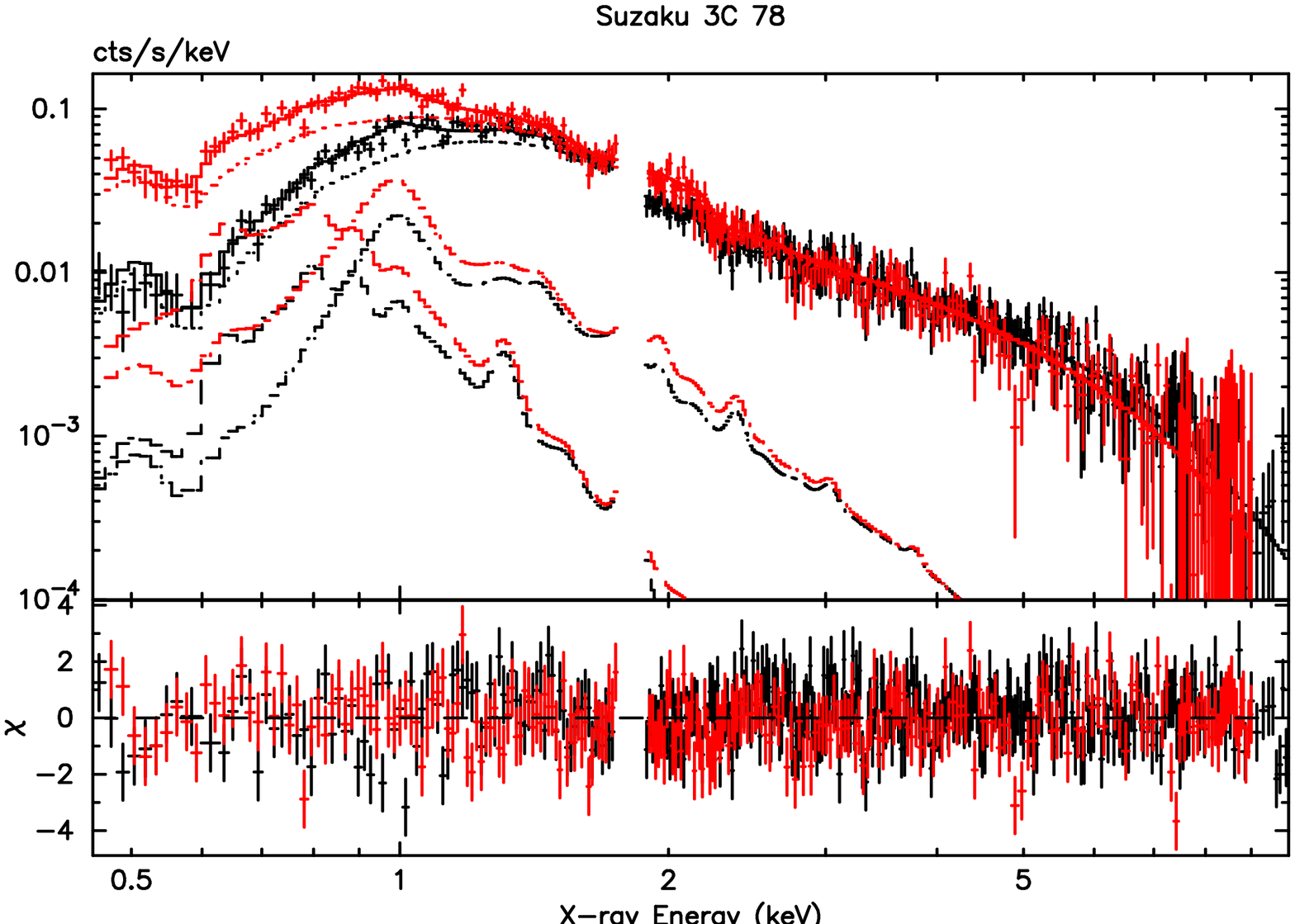}
\vspace{0.5cm}
\caption{{\em Suzaku} spectra of \pks and \tc. The black, red, and green symbols are XIS-F, XIS-B, and HXD-PIN spectra, respectively. The solid line represents the best-fit total model, while the dashed and dotted lines are the {\tt apec} and power-law model components, respectively. The bottom panels show the residuals in units of $\sigma$.}
\label{xspec}
\end{center}
\end{figure}

\begin{figure}[!t]
\begin{center}
\vspace{-1cm}
\includegraphics[scale=0.32]{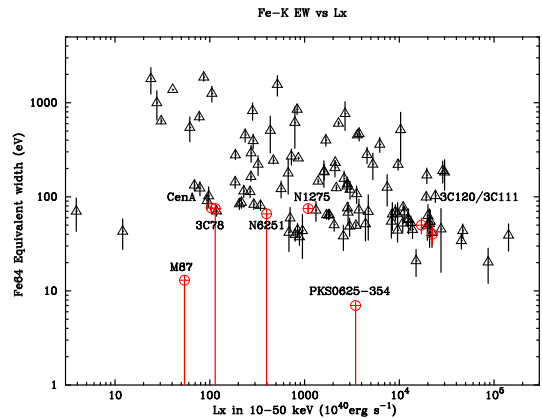}
\vspace{0.5cm}
\caption{Fe-K line EW plotted against the X-ray luminosity for our sample of radio galaxies (red circles) and Seyfert galaxies (black triangles) analyzed by \citet{fuk11a}. The data points with only the lower error bar represent upper limits.}
\label{fek}
\end{center}
\end{figure}

Figure \ref{xspec} shows {\em Suzaku} X-ray spectra of \pks and \tc.
Quality of X-ray spectra are better than ever for both objects.
We fitted their spectra by one or two plasma model plus power-law
model.
The former represents a soft thermal emission associated with host
galaxies.
The spectra are well fitted by this modeling, and the power-law photon
index is 2.25$\pm$0.02 and 2.32$\pm$0.04 for \pks and \tc,
respectively.
This value is relatively larger for Seyfert galaxies whose X-ray
emission is dominated by disk/corona emission.
Fe-K line is not detected for both; an upper limit of equivalent width (EW)
of Fe-K line is 7 eV and 75 eV for \pks and \tc, respectively.
X-ray time variation during one {\em Suzaku} observation is weak or
little.
We compared X-ray peoperties with those of Seyfert galaxies, together
with other GeV-emitting radio galaxies.
Fe-K line EWs of \pks, \tc, M\,87, NGC\,6251 are smaller tha those of
typical Seyfert galaxies as shown in figure \ref{fek}.
Fe-K line is emitted when the X-ray emission from the central
disk/corona region is reflected by the dust torus with a large
reflection angle.
Therefore, a weak or no Fe-K line indicates that the X-rays are
not a disk/corona emission but likely  a beamed jet emission.
X-ray luminosity of \pks is higher than that of typical Seyfert
galaxies with a smilar [O III] luminosity.
Combined with studies of other X-ray properties, such as spectral
index, variability, X-ray emission of low excitation radio galaxies
(LERG), which are considered to have a low mass accretion rate, is
likely to be a jet emission.

\section{GeV Gamma-ray results and SED}

We analyzed {\em Fermi} LAT 5 years data of \pks and \tc.
For the analysis, LAT Science Tools version v9r32p5 was utilized 
with the {\tt P7REP\_SOURCE\_V15} Instrument Response Functions
(IRFs). 
Both radio galaxies are clearly visible in the 0.2 to 300\,GeV LAT 
counts maps. 
We extracted the data within a 12$\times$12\,deg$^2$ rectangular
region centered on each object.  The binned likelihood fitting with
the {\tt gtlike} tool was performed.  The field background point
sources within 14.5$^{\circ}$ from each source, listed in the LAT 2
year catalog \citep{nol12}, were included.
The standard LAT Galactic emission model was used ({\tt
gll\_iem\_v05.fits}) and the isotropic diffuse gamma-ray background
and the instrumental residual background were represented as a uniform
background ({\tt iso\_source\_v05.txt}). 
A likelihood analysis was performed with the energy information binned
logarithmically in 30 bins in the 0.2--300\,GeV band, and the spatial
information binned with 0.15$\times$0.15\,deg$^2$ bin size. 

\begin{figure}[!tb]
\includegraphics[scale=0.32]{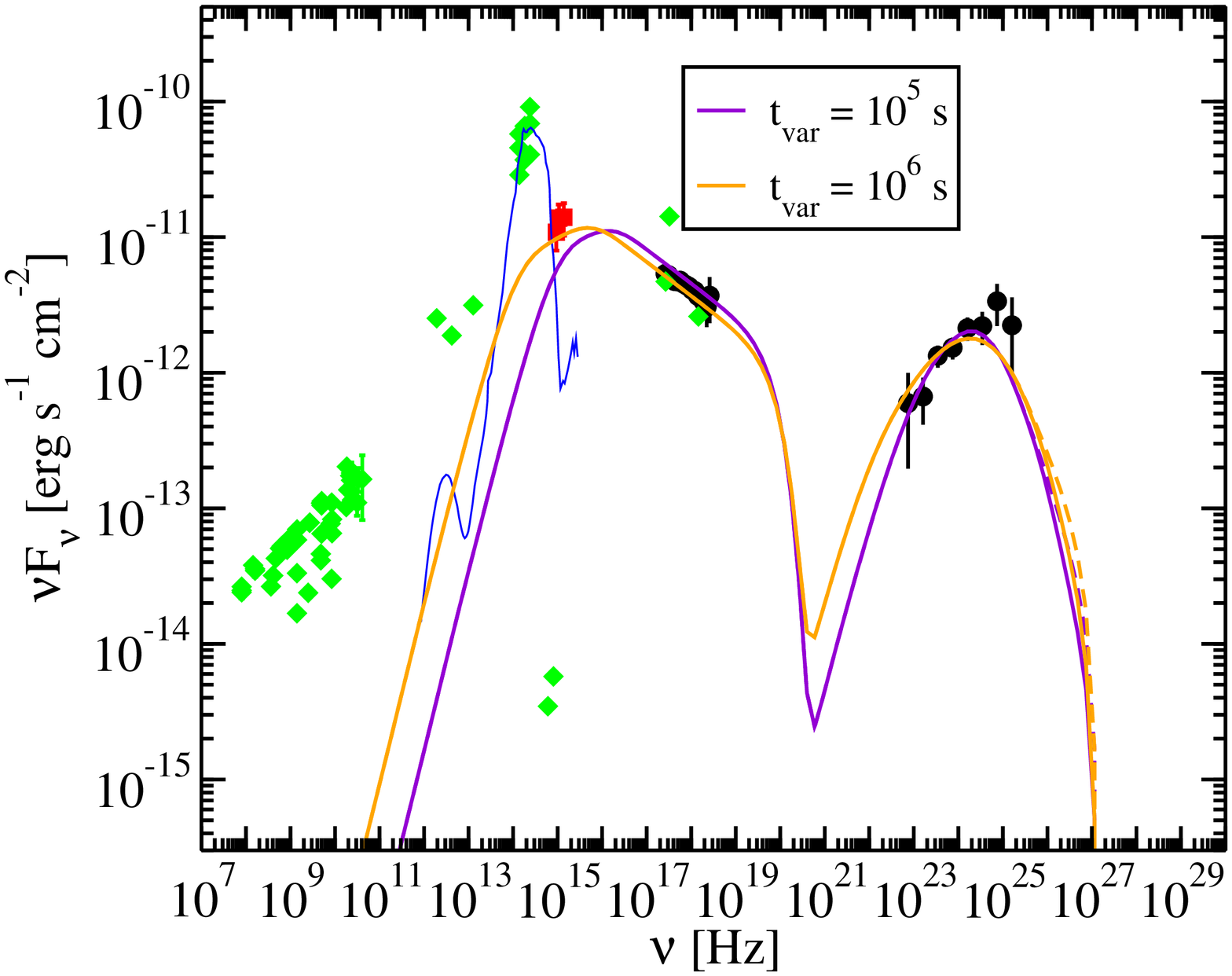}
\begin{center}
\caption{SEDs of \pks. Black circles indicate the {\em Suzaku} X-ray and {\em Fermi}-LAT $\gamma$-ray data presented in this paper, green diamonds are archival data. The thick curves denote the synchrotron/SSC model fits with two different variability timescales, as given in the legend. The solid curves include $\gamma\gamma$ absorption with the EBL model of \citet{fin10}, while the dashed curves do not. The thin blue curves are the elliptical galaxy template.}
\label{sed1}
\end{center}

\includegraphics[scale=0.32]{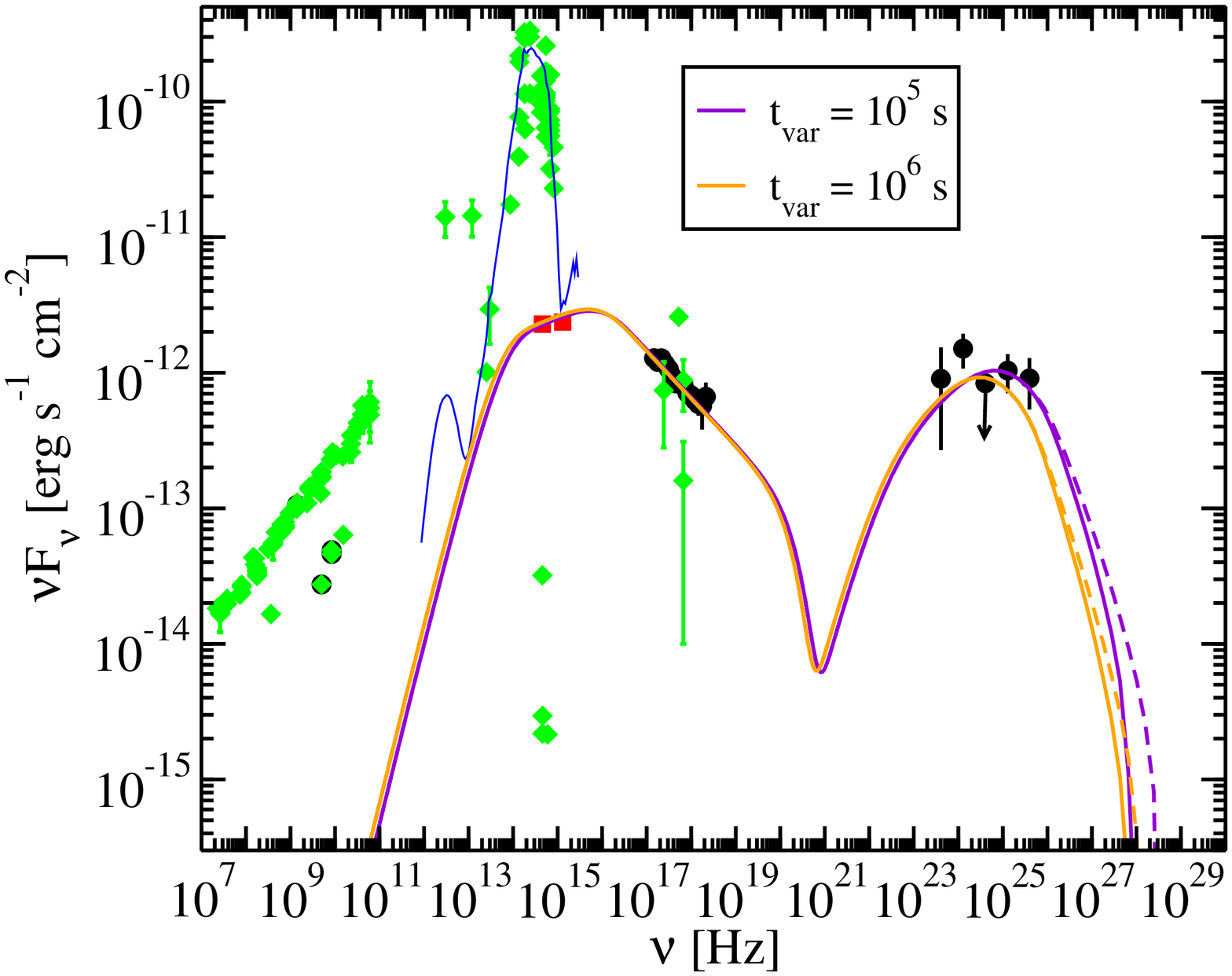}
\begin{center}
\caption{SEDs of \tc. Same as figure \ref{sed1}.}
\label{sed2}
\end{center}
\end{figure}

\begin{figure}{it}
\begin{center}
\includegraphics[scale=0.32]{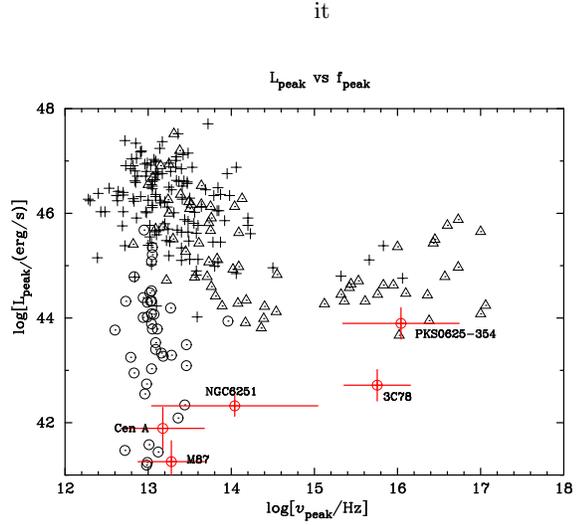}
\vspace{0.5cm}
\caption{Relation between synchrotron peak frequencies and peak luminosities of \pks\ and \tc, together with other sources from our sample of radio galaxies (red {\bf circles}). For a comparison, radio galaxies, BL Lacs, and FSRQs from \citet{mey11} are also plotted (black circles, triangles, and crosses, respectively).}
\label{fplp}
\end{center}
\end{figure}

Gev gamma-ray spectra of both galaxies show a flat power-law with a
photon index of 1.72$\pm$0.06 and 2.01$\pm$0.16 for \pks and \tc,
respectively.
Studies of time variability show  a flare-like event for \pks and no
significant variation for \tc.
Figure \ref{sed1} and \ref{sed2} shows a SED of both galaxies, based on our {\em
Suzaku} and {\em Fermi} data and other available data.
SEDs are well modellded by the one-zone synchrotron self-Compton model 
from \citet{fin08}.
Compared with other GeV-emitting radio galaxies whose results were
also all done by the same model of \citet{fin08}, lower bulk Lorentz
factors of 2--6 are preferred when compared to those of typical blazars.
An unique property of \pks and \tc is a higher breaking energy of
electron spectrum.
This is attributed to higher SED-peaking energies of both galaxies (figure
\ref{sed1}, \ref{sed2}).
Considering this property, we plot the Synchrotron peak luminosity
against the Synchrotron peaking frequency as shown in figure
\ref{fplp}, where most of other data of blazars and radio galaxies are
taken from \citet{mey11}.
For this plot, \citet{mey11} states the high-E peaked objects are only
the most aligned jet objects with radiatively inefficient accretion
and decelerating weak jet.
However, \pks and \tc are outliers of this model, and thus they are at
odds with the FR-I/BL Lac unification.

\clearpage
\begin{thebibliography}{99}

\bibitem[Abdo et al.(2010)]{abd10b} Abdo, A.~A., Ackermann, M., Ajello, M., et al.\ 2010b, Science, 328, 725
\bibitem[Finke et al.(2008)]{fin08} Finke, J.~D., Dermer, C.~D., B\"ottcher, M.\ 2008, ApJ, 686, 181 
\bibitem[Fukazawa et al.(2011)]{fuk11a} Fukazawa, Y., Hiragi, K., Mizuno, M., et al.\ 2011a, ApJ, 727, 19 
\bibitem[Fukazawa et al.(2015)]{fuk14} Fukazawa, Y., Finke, J., Stawarz, {\L}., et al.\ 2015, ApJ, 798, 74 
\bibitem[Meyer et al.(2011)]{mey11} Meyer, E.~T., Fossati, G., Georganopoulos, M., \& Lister, M.~L.\ 2011, ApJ, 740, 98 
\bibitem[Nolan et al.(2012)]{nol12} Nolan, P.~L., Abdo, A.~A., Ackermann, M., et al.\ 2012, ApJS, 199, 31 

\end{thebibliography}
\end{document}